\documentclass[reprint,prl,preprintnumbers,superscriptaddress,amsmath,amssymb]{revtex4-1}
\usepackage{hyperref}
\usepackage{color}
\usepackage{float}
\usepackage{natbib}
\usepackage{graphicx}% Include Figure files
\usepackage{dcolumn}% Align table columns on decimal point
\usepackage{bm}% bold math
\usepackage{siunitx} % to write um so that the $\mu$ doesn't appear italic, e.g. "`The average time of this event is \SI{50}{\micro\second}."'
\newcommand\tab[1][0.5cm]{\hspace*{#1}}
\begin{document}

\bibliographystyle{unsrtnat}
\title{\large{Large, valley-exclusive Bloch-Siegert shift in monolayer WS$_2$}}

\vspace{0.5cm}
\author{Edbert J. Sie}
	\affiliation{Department of Physics, Massachusetts Institute of Technology, Cambridge, MA 02139, USA}
\author{Chun Hung Lui}
		\affiliation{Department of Phyiscs and Astronomy, University of California, Riverside, California 92521, USA}		
\author{Yi-Hsien Lee}
		\affiliation{Materials Science and Engineering, National Tsing-Hua University, Hsinchu 30013, Taiwan}
\author{Liang Fu}
		\affiliation{Department of Physics, Massachusetts Institute of Technology, Cambridge, MA 02139, USA}
\author{Jing Kong}
		\affiliation{Department of Electrical Engineering and Computer Science, Massachusetts Institute of Technology, Cambridge, MA 02139, USA}
\author{Nuh Gedik}
	\altaffiliation{gedik@mit.edu}
	\affiliation{Department of Physics, Massachusetts Institute of Technology, Cambridge, MA 02139, USA}
\vspace{0.5cm}

\maketitle

\noindent
\textbf{\small{Coherent light-matter interaction can be used to manipulate the energy levels of atoms, molecules and solids. When light with frequency $\omega$ is detuned away from a resonance $\omega_0$, repulsion between the photon-dressed (Floquet) states can lead to a shift of energy resonance. The dominant effect is the optical Stark shift ($\propto 1/(\omega_0-\omega)$), but there is an additional contribution from the so-called Bloch-Siegert shift ($\propto 1/(\omega_0+\omega)$). Although it is common in atoms and molecules, the observation of Bloch-Siegert shift in solids has so far been limited only to artificial atoms since the shifts were small ($<$1 $\mu$eV) and inseparable from the optical Stark shift. Here we observe an exceptionally large Bloch-Siegert shift ($\sim$10 meV) in monolayer WS$_2$ under infrared optical driving by virtue of the strong light-matter interaction in this system. Moreover, we can disentangle the Bloch-Siegert shift entirely from the optical Stark shift, because the two effects are found to obey opposite selection rules at different valleys. By controlling the light helicity, we can confine the Bloch-Siegert shift to occur only at one valley, and the optical Stark shift at the other valley. Such a valley-exclusive Bloch-Siegert shift allows for enhanced control over the valleytronic properties in two-dimensional materials, and offers a new avenue to explore quantum optics in solids.}}

\begin{figure}[h!]
	\includegraphics[width=0.45\textwidth]{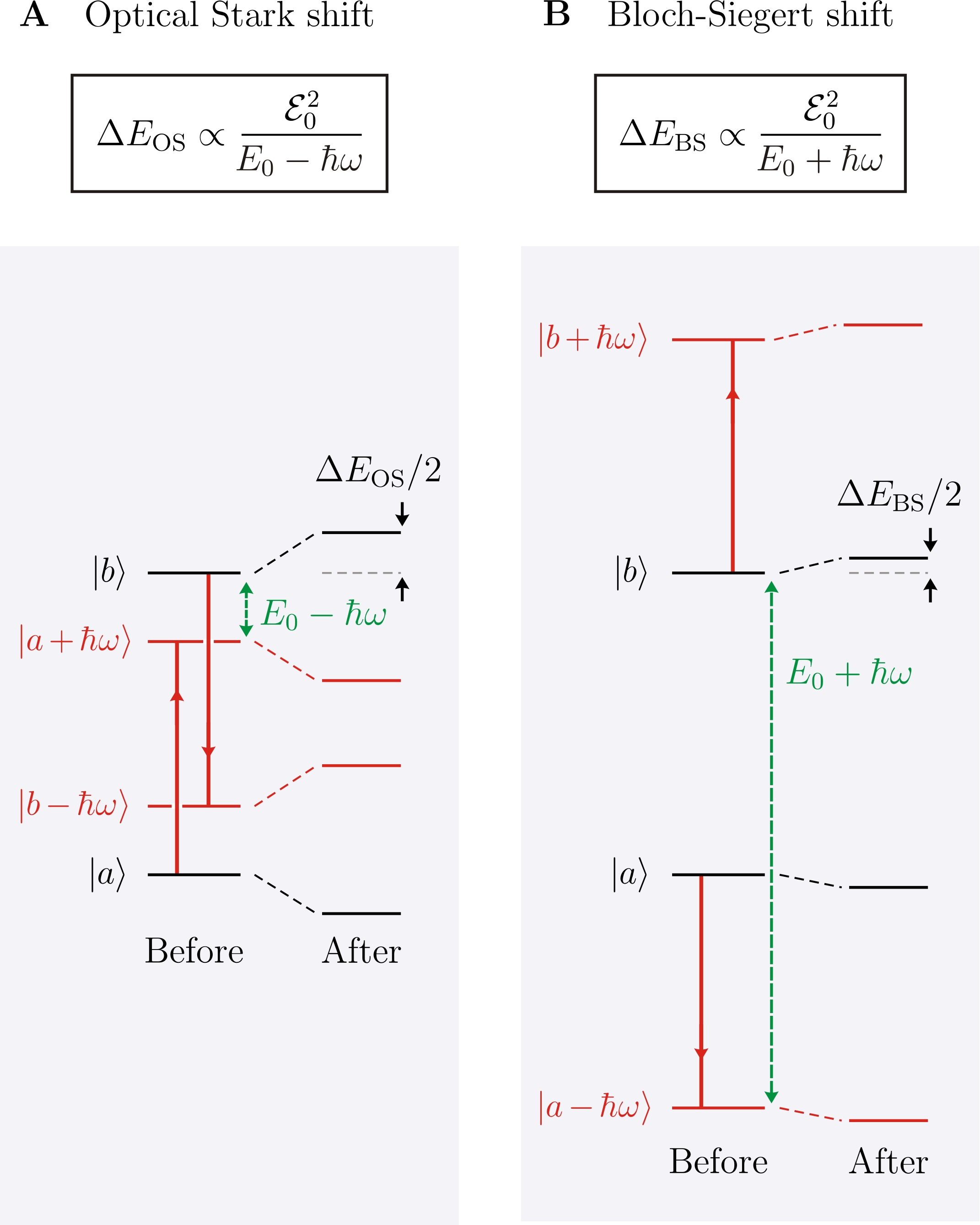}
	\caption{\textbf{Comparison of the optical Stark shift and the Bloch-Siegert shift in a two-level system.}  (\textbf{A}) Energy diagram for optical Stark (OS) shift.  $\left|a\right\rangle$ and $\left|b\right\rangle$ denote the two original states with resonance energy $E_0$ before they are optically driven. $\left|a+\hbar\omega\right\rangle$ and $\left|b-\hbar\omega\right\rangle$ are photon-dressed (Floquet) states driven by the co-rotating optical field. Hybridization with these Floquet states causes the resonance energy to blueshift by $\Delta E_{\text{OS}}$, which is proportional to the light intensity ($\mathcal{E}_0^2$) and inversely proportional to their energy separation ($E_0 -\hbar\omega$). (\textbf{B}) Energy diagram for Bloch-Siegert (BS) shift. $\left|a-\hbar\omega\right\rangle$ and $\left|b+\hbar\omega\right\rangle$ are two different Floquet states driven by the counter-rotating optical field. Hybridization with these Floquet states causes the Bloch-Siegert shift, with magnitude $\Delta E_{\text{BS}}$ inversely proportional to their energy separation ($E_0 + \hbar\omega$).}
	\label{fig:Fig1}
\end{figure}

The fundamental interaction between light and matter can be understood within the framework of a two-level system \cite{Cohen98, Shirley65}. When driven by off-resonant light $\hbar\omega < \hbar\omega_0 (= E_0)$, there are two pairs of photon-dressed (Floquet) states which contribute to the state repulsion with the original states -- one pair between the original states (Fig. 1A) and the other pair outside the original states (Fig. 1B). The former case leads to a shift of transition energy called the optical Stark (OS) shift, which increases linearly with the light intensity ($\mathcal{E}_0^2$) and inversely with the detuning energy, $\Delta E_{\text{OS}} \propto \mathcal{E}_0^2/(E_0-\hbar\omega)$ \cite{Autler55}. The latter case also leads to a shift, called the Bloch-Siegert (BS) shift, but it has a different energy dependence, $\Delta E_{\text{BS}} \propto \mathcal{E}_0^2/(E_0+\hbar\omega)$ \cite{Bloch40}. Although the Bloch-Siegert shift is negligible at small detuning, it can become comparable, and serves as an important correction, to the optical Stark shift at large detuning.

\begin{figure*}[t]
	\includegraphics[width=0.97\textwidth]{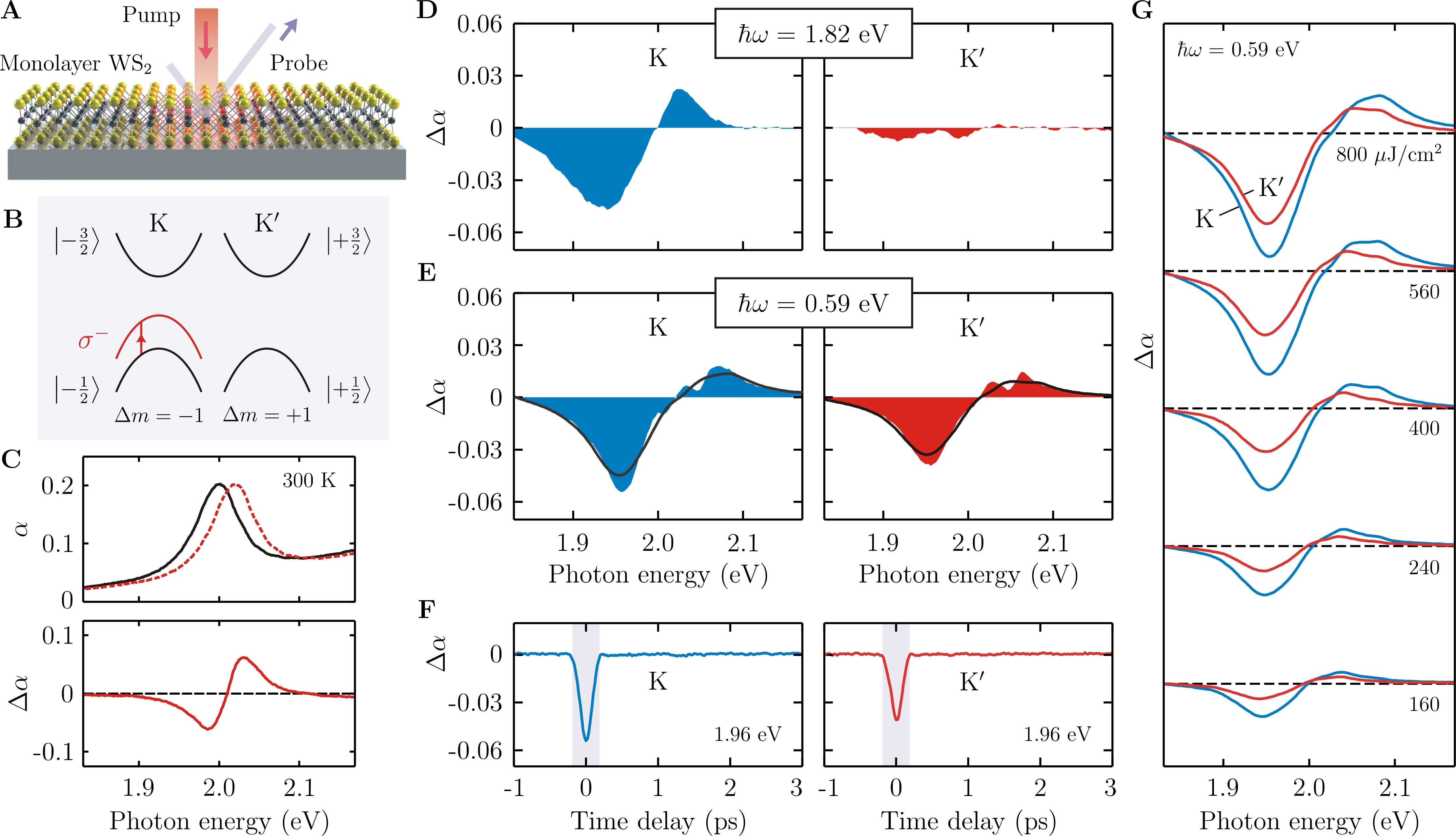}
	\caption{\textbf{Observation of valley-exclusive Bloch-Siegert shift in monolayer WS$_2$.} (\textbf{A}) Illustration of the pump-probe experiment. We pump the monolayer WS$_2$ sample with strong infrared pulses, and measure the pump-induced change of reflection with broadband probe pulses.  (\textbf{B}) The K and K$^{\prime}$ valleys in monolayer WS$_2$. Optical pumping with left-handed circular polarization ($\sigma^-$) couples only to the K valley, and not the K$^{\prime}$ valley, unless the counter-rotating field is taken into account.  (\textbf{C}) Measured A exciton absorption spectrum ($\alpha$, black curve in the top panel) of monolayer WS$_2$ in equilibrium at room temperature. The dashed curve represents the shifted resonance (simulated) under red-detuned optical pumping. This shift produces a differential curve in the absorption change ($\Delta\alpha$, red curve in the bottom panel).  (\textbf{D} and \textbf{E}) The $\Delta\alpha$ spectra under zero-delay optical pumping at photon energy 1.82 and 0.59 eV. By using probe pulses with $\sigma^-$ ($\sigma^+$) polarization, we can selectively measure $\Delta\alpha$ at the K (K$^{\prime}$) valley, as shown in the left (right) column. The black curves in (E) are smoothened curves to average out the modulations, as mentioned in the text.  (\textbf{F}) Time trace of $\Delta\alpha$ spectra in (E) measured at probe energy of 1.96 eV. The induced energy shift is observed only at zero time delay.  (\textbf{G}) The zero-delay $\Delta\alpha$ spectra of the K valley (blue curves) and K$^{\prime}$ valley (red curves) under different incident pump fluence ($\mathcal{F}$ = 160$-$\SI{800}{\micro\J}/cm$^2$). The pump photon energy is 0.59 eV. The spectra are vertically displaced for clarity.}
	\label{fig:Fig2}
\end{figure*}

The Bloch-Siegert shift has played an important role in atomic physics, notably for its manifestation as the Lamb shift in quantum electrodynamics \cite{Lamb47, Bethe47} and its contribution to the trapping potential for cold atoms \cite{Grimm00}. In condensed matter physics, however, the Bloch-Siegert shift is a very rare finding because so far the shifts were tiny and can only be revealed indirectly by subtracting the dominant optical Stark shift with sophisticated modeling \cite{Forn10, Tuorila10, Niemczyk10}. To elucidate the detailed characteristic of Bloch-Siegert shift, it is necessary to separate the two effects. Now, considering that they are time-reversed partners of each other -- the optical Stark shift arises from co-rotating field and the Bloch-Siegert shift from counter-rotating field -- it is theoretically possible to separate them under stimulation that breaks time-reversal symmetry. Experimental realization of this novel scheme has, however, not been demonstrated thus far. 

We report on the observation of an unprecedentedly large Bloch-Siegert shift ($\Delta E_{BS} \sim$ 10 meV), which can be entirely separated from the optical Stark shift. Such a large and exclusive Bloch-Siegert shift is realized in a monolayer of transition-metal dichalcogenide (TMD) tungsten disulfide (WS$_2$). This is possible because this material system possesses two distinctive features. First, it exhibits strong light-exciton interaction at the two time-reversed valleys (K, K$^{\prime}$) in the Brillouin zone (Fig. 2A-C) \cite{Mak10, Chernikov14, Liu14}. Secondly, the two valleys possess finite and opposite Berry curvatures due to the lack of inversion symmetry, giving rise to distinct optical selection rules and related valleytronic properties \cite{Xiao12, Mak12, Zeng12, Cao12, Wu13, Xu14, Mak14, Rivera16, Schaibley16}. That is, the optical transition at the K (K$^{\prime}$) valley is coupled exclusively to left-handed $\sigma^-$ (right-handed $\sigma^+$) circularly polarized light. This leads to a unique material platform that allows us to separate the Bloch-Siegert shift from the optical Stark shift by using circularly polarized light.

We employ femtosecond pump-probe absorption spectroscopy in our experiment (Fig. 2A). We pump a monolayer of WS$_2$ with intense $\sigma^-$ infrared light pulses and probe the energy shift at the K (K$^{\prime}$) valley with $\sigma^-$ ($\sigma^+$) visible light pulses (see Supplementary Materials). A blueshift of the exciton absorption peak ($\alpha$) is manifested as a differential curve in the absorption change $\Delta\alpha$ (Fig. 2C). From this, we can deduce the magnitude of the energy shift at both valleys (see Supplementary Materials). Previously, transient absorption with visible pumping has been used to study the optical Stark effect in monolayer TMDs \cite{Sie15, Kim14}. Here, by pumping with infrared light, we reveal Bloch-Siegert shift in WS$_2$ for the first time.

Figure 2(D-G) display our results at the K valley (blue curve) and K$^{\prime}$ valley (red curve). For comparison, we first show the $\Delta\alpha$ spectra under zero-delay pumping at $\hbar\omega =$ 1.82 eV (Fig. 2D).  For this small detuning energy, only the K valley shows an appreciable $\Delta\alpha$ signal. This signal arises from the optical Stark shift, which occurs exclusively at the K valley \cite{Sie15, Kim14}. The K$^{\prime}$ valley exhibits only very weak (but observable) signal. However, as we lower the pumping photon energy to 0.59 eV, the signal at the K$^{\prime}$ valley becomes comparable to the signal at the K valley (Fig. 2E). This observation indicates a pronounced energy blueshift at the K$^{\prime}$ valley, a phenomenon that apparently violates the well-established valley selection rules in monolayer TMDs.  Some minor modulation features also appear in the $\Delta\alpha$ spectrum, but they are irrelevant to the current study and will be explored in a separate work. We average out these modulations by slightly smoothening the curves (Supplementary Materials). We have further examined the signals at different pump-probe time delay. The $\Delta\alpha$ signals at both valleys emerge only at zero time delay, with very similar temporal profiles to the 160 fs duration of the pump pulses (Fig. 2F). These results indicate the coherent nature of the energy shift, and also exclude the effect from intervalley scattering of possible excited carriers that typically occurs in the picosecond time scale \cite{Mai14, Mai14B, Sie15B}. 

To investigate the underlying mechanism of the anomalous energy shift at K$^{\prime}$ valley, we have measured the zero-delay  spectra for both valleys at various pump photon energies ($\hbar\omega =$ 0.59, 0.69, 0.89, 0.98 eV) and different pump fluences ($\mathcal{F}$ = 30$-$\SI{800}{\micro\J}/cm$^2$). Here we display the fluence-dependent spectra for pump photon energy $\hbar\omega =$ 0.59 eV (Fig. 2G), while the remaining spectra are presented in the Supplementary Materials. The $\Delta\alpha$ spectra at both valleys are found to grow with increasing pump fluence. For a more quantitative analysis, we have extracted the energy shift from each spectrum, and plotted it as a function of $\mathcal{F}/(E_0-\hbar\omega)$ (Fig. 3A). The shift at K valley exhibits an excellent linear dependence regardless of the different pump photon energies (closed symbols), indicating that it arises from the optical Stark effect.  The shift at K$^{\prime}$ valley, however, spreads out with no rigorous linear dependence (open symbols). Such a contrasting behavior indicates that the K$^{\prime}$-valley shift does not arise from the optical Stark effect. In Figure 3B, we replot the K$^{\prime}$-valley shift as a function of $\mathcal{F}/(E_0+\hbar\omega)$ with the same axes scales. Remarkably, the data now exhibit an excellent linear dependence. Moreover, the slope of the K$^{\prime}$-valley shift in this new plot is identical to the slope of the K-valley shift in Fig. 3A. This observation strongly suggests that the K$^{\prime}$-valley shift arises from the Bloch-Siegert effect.

\begin{figure*}[t]
	\includegraphics[width=0.75\textwidth]{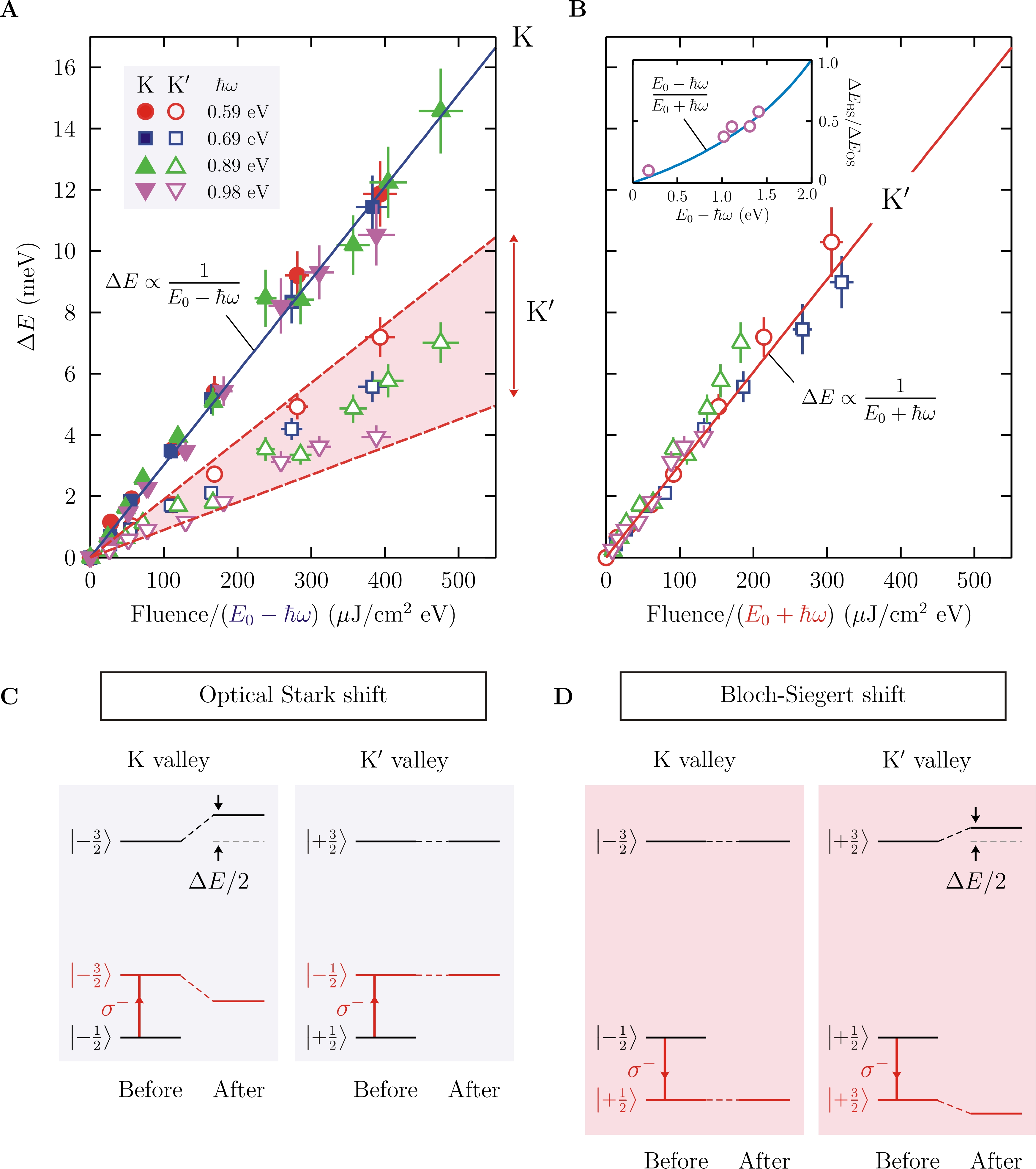}
	\caption{\textbf{Fluence and detuning dependences of the Bloch-Siegert shift.}  (\textbf{A}) Energy shifts at the K and K$^{\prime}$ valleys (closed and open symbols) as a function of fluence $\mathcal{F}/(E_0-\hbar\omega)$. The data are extracted from the $\Delta\alpha$ spectra in Fig. 2 and Fig. S1 in the Supplementary Materials. The K-valley shift exhibits a rigorous linear dependence (solid blue line), but the K$^{\prime}$-valley shift spreads out (red region).  (\textbf{B}) Energy shift at K$^{\prime}$ valley as in (A), but plotted as a function of fluence $\mathcal{F}/(E_0+\hbar\omega)$. The linear dependence becomes obvious. The top-left inset shows the predicted ratio $\Delta E_{BS}/\Delta E_{OS}$ (blue line) when the detuning energy ($E_0-\hbar\omega$) increases from zero to the resonant energy ($E_0 =$ 2 eV for monolayer WS$_2$). The open circles are our averaged experimental data obtained from Fig. 2 and Fig. S1.  (\textbf{C}) Energy diagram before and after optical pumping for the optical Stark shift, which occurs only at the K valley.  (\textbf{D}) Energy diagram for the Bloch-Siegert shift, which occurs only at the K$^{\prime}$ valley.}
	\label{fig:Fig3}
\end{figure*}

Our finding can be verified quantitatively by using either a semi-classical theory or a fully quantum-mechanical theory \cite{Crisp91} (see the details of both treatments in the Supplementary Materials). As we probe only the lowest-energy exciton state (1s), which shows similar properties as those of hydrogen atoms, it is appropriate and sufficient to use a simple two-level framework, as shown in earlier studies \cite{Chernikov14, Mak12, Sie15, Kim14}. In our semi-classical analysis, we treat the ground state and the 1s exciton state as the two-level system ($\left|a\right>$ and $\left|b\right>$) with a resonance energy $E_0$, driven by a classical electromagnetic wave with amplitude $\mathcal{E}_0$ and frequency $\omega$. We use a left-circularly polarized pump beam $\vec{\mathcal{E}} (t) = \mathcal{E}_0 \left(\cos(kz-\omega t) \hat{x}+ \sin(kz-\omega t) \hat{y}\right)$, polarized along the $xy$-plane of the monolayer sample ($z=0$), which can also be expressed as:
\begin{equation}
\vec{\mathcal{E}}(t) = \frac{1}{2}\mathcal{E}_0\left[(\hat{x}-i\hat{y})e^{-i\omega t} + (\hat{x}+i\hat{y})e^{i\omega t}\right]
\end{equation}
where the field is decomposed into two terms based on their time-evolution. The interaction Hamiltonian can then be expressed as
\begin{eqnarray}
H_{ab} &=& \left\langle b\left|e\vec{\mathcal{E}}\cdot \vec{r}\right|a\right\rangle\\
       &=& \frac{1}{2} e\mathcal{E}_0\left\langle b\left|(\vec{x}-i\vec{y})e^{-i\omega t} + (\vec{x}+i\vec{y})e^{i\omega t}\right|a\right\rangle
\end{eqnarray}
Here we can see that the first term $(\vec{x}-i\vec{y})e^{-i\omega t}$ induces a transition with $\Delta m = -1$ (co-rotating field), and the second term $(\vec{x}+i\vec{y})e^{i\omega t}$ induces a transition with $\Delta m = +1$ (counter-rotating field). Due to the unique valley selection rules in monolayer WS$_2$, these two terms are thus coupled exclusively to the K ($\Delta m = -1$) and K$^{\prime}$ ($\Delta m = +1$) valleys, respectively (Fig. 3C-D), with their valley-specific interactions as:
\begin{eqnarray}
H_{ab} (\text{K}) = e^{-i\omega t} \mu_\text{K}\mathcal{E}_0 /2\\
H_{ab} (\text{K$^{\prime}$}) = e^{i\omega t} \mu_{\text{K$^{\prime}$}}\mathcal{E}_0 /2
\end{eqnarray}
Here $\mu_K$ and $\mu_{K^{\prime}}$ are the dipole matrix elements at the K and K$^{\prime}$ valley, respectively, and they have equal magnitudes $\mu = \left|\mu_K\right| = \left|\mu_{K^{\prime}}\right|$. However, they are associated with opposite time-evolution factors, which lead to a more general theory of valley selection rules in monolayer TMDs. Under resonant absorption condition ($\omega = \omega_0$), the left-circularly polarized light couples only to the K-valley. But under off-resonance condition ($\omega < \omega_0$), the coupling to the K$^{\prime}$-valley can become significant through the time-reversed process, giving rise to noticeable energy shift. The induced energy shifts at the respective valleys can be evaluated by the time-dependent perturbation theory as:
\begin{eqnarray}
\Delta E_{\text{K}} = \frac{\mu^2\mathcal{E}_0^2}{2} \frac{1}{E_0 - \hbar\omega}\\
\Delta E_{\text{K}^{\prime}} = \frac{\mu^2\mathcal{E}_0^2}{2} \frac{1}{E_0 + \hbar\omega}
\end{eqnarray}
The two energy shifts have different energy dependence, from which we can readily identify $\Delta E_\text{K}$ to be the optical Stark shift and $\Delta E_\text{K}^{\prime}$ the Bloch-Siegert shift. When plotted as a function of their respective energy denominator ($E_0-\hbar\omega$, or $E_0+\hbar\omega$), both shifts exhibit an identical slope. The prediction of common slope and opposite valley indices agrees well with our experimental observation (Fig. 3A-B).  From our data, we can deduce the dipole matrix elements to be $\mu =$ 55 Debye, in excellent agreement with previous measurements \cite{Sie15}. In addition, the ratio between $\Delta E_\text{K}^{\prime}$ and $\Delta E_\text{K}$ is predicted to be $(E_0-\hbar\omega)/(E_0+\hbar\omega)$, the same as $\Delta E_{BS}/\Delta E_{BS}$ for a generic two-level system. By plotting the average shift ratio measured for each pump photon energies, we find a good agreement between our experiment and theory (see inset of Fig. 3B).

The physics of this valley-exclusive energy shift can be illustrated in the energy diagrams shown in Fig. 3C-D. The co-rotating field generates a Floquet state $\hbar\omega$ above the ground state in both valleys, with energy separation $E_0-\hbar\omega$ from the excited state. Due to the matching condition of angular momentum, repulsion between the Floquet state and the excited state only occurs at the K valley, giving rise to the ordinary optical Stark shift (Fig. 3C). On the other hand, the counter-rotating field generates a Floquet state $\hbar\omega$ below the ground state, with energy separation $E_0+\hbar\omega$ from the excited state (Fig. 3D). The matching condition of angular momentum forbids the level repulsion at the K valley but allows it at the K$^{\prime}$ valley. This gives rise to the Bloch-Siegert shift at the opposite (K$^{\prime}$) valley. In other words, the left-circularly polarized light can be understood as stimulating the $\sigma^-$ absorption ($\Delta m = -1$) and $\sigma^-$ emission ($\Delta  = +1$) processes at K and K$^{\prime}$ valleys, respectively. This unique mechanism demonstrates that the left circularly polarized light can in fact couple to both valleys distinctively, thus establishing a new concept of valley selection rules.

In summary, we have presented the observation of a large, valley-exclusive Bloch-Siegert shift in monolayer WS$_2$. This shift exhibits the opposite valley selection rules from the ordinary optical Stark effect, which allows us to completely separate the two effects. This is possible because, as time-reversed partners, the two effects share similar relationship with the two time-reversed valleys in monolayer TMDs, which can be disentangled under circularly polarized light that breaks time-reversal symmetry. Our finding reveals more general valley selection rules and may lead to enhanced control over the valleytronic properties of Dirac materials such as graphene, TMDs, and Weyl semimetals \cite{Kundu16}.

\vspace{0.5cm}
\noindent
\textbf{Acknowledgments}
\newline
\small{N.G. and E.J.S are grateful to Martin Zwierlein and Junichiro Kono for the stimulating discussions. We thank Qiong Ma and Yaqing Bie for the assistance during absorption measurement. N.G. and E.J.S acknowledge funding supports from the U.S. Department of Energy, BES DMSE (experimental setup and data acquisition), and from the Gordon and Betty Moore Foundation's EPiQS Initiative grant GBMF4540 (data analysis and manuscript writing). J.K. and L.F. acknowledge funding support from NSF Science and Technology Center for Integrated Quantum Materials grant DMR-1231319 (material growth and theory). Y-H.L. acknowledges funding supports from the Asian Office of Aerospace Research and Development/Office of Naval Research Global grant FA2386-16-1-4009 and Ministry of Science and Technology Taiwan, grants 105-2112-M-007-032-MY3 and 105-2119-M-007-027) (material growth).}

\vspace{0.5cm}
\noindent
\textbf{Supplementary Materials}
\newline
\href{http://www.sciencemag.org/content/355/6329/1066/suppl/DC1}{\color{blue} www.sciencemag.org/content/355/6329/1066/suppl/DC1}
\newline
\small{Materials and Method\\
Supplementary Text:\\
\tab 1. Supplementary data\\
\tab 2. Extraction of the energy shift from $\Delta\alpha$ spectra\\
\tab 2. Semi-classical derivation of Bloch-Siegert shift\\
\tab 3. Quantum-mechanical derivation of Bloch-Siegert shift\\
Figures S1-S3\\
References \cite{Lee12, Lee13, Chemla89, Bransden03, Jaynes63}}

\end{document}